\documentstyle[aps,prb,psfig]{revtex}

\begin{document}
%\begin{minipage}[t]{16cm}
%\onecolumn
  \draft
%  \narrowtext
\title{Variational quantum Monte Carlo ground state of GaAs}
\author{H.~Eckstein and W.~Schattke}
\address{Institut f\"ur Theoretische Physik, % \\
      Universit\"at Kiel, % \\
      D-24098 Kiel, Germany}
\author{M.~Reigrotzki and R.~Redmer}
\address{ Fachbereich Physik, % \\
      Universit\"at Rostock 
      Universit\"atsplatz 3, % \\
      D-18051 Rostock, Germany}
\date{Received 18 March 1996}

\maketitle
\begin{abstract}
Variational quantum Monte Carlo calculations are reported for the bulk GaAs
semiconductor in order to present values for the ground-state energy, 
the lattice constant, the bulk modulus, and some derived properties. 
The statistical accuracy is significantly higher 
than the remaining differences to the experimental values, 
especially for the total-energy upper bound. 
The agreement with experiment is satisfactory. 
The results are also compared with those of density 
functional calculations. The accuracy of our results is comparable
to the best of those calculations.
\end{abstract}
% \vspace{1cm}
%\end{minipage}
%\\
%\begin{minipage}[t]{20cm}
% \twocolumn
\section{Introduction}
In comparison with the widely accepted and extensively applied local density 
functional technique (LDA), the quantum Monte Carlo methods (QMC) represent 
rather new developments for the ground-state determination of solids.
\cite{McMillan,CeperleyCK77,CeperleyAlder80,CeperleyAlder87,Fahy,SugiyamaAlder,EcksteinSchattke,CancioChang,Foulkes} 
Especially the variational quantum Monte Carlo method (VQMC) is attractive, 
because the quality of the approximation is controlled 
by the property of yielding an upper bound, i.e., by the simple rule: 
the lower the energy the better the wave function. The statistical error 
of an expectation value can be safely estimated. 
It behaves favorably near an eigenvalue where the variance should vanish. 
Thus this method is really {\it ab initio}, not relying on any uncontrolled 
quantity, such as an unknown correlation functional in the case of LDA. 
This aspect seems to be actually important, because the detailed results 
of the density functional theory (DFT) in fact depend sensitively on the choice of this functional.\cite{Garcia,Ortiz,Filippi,Causa} 
Furthermore, in the past the role of Monte Carlo simulations for 
homogeneous systems was to determine the correlation energy to be adapted 
as functional for inhomogeneous LDA calculations.\cite{CeperleyAlder80} 
Thus today's QMC calculations on inhomogeneous systems may also be valued 
by the DFT as it may lead to an improved functional.

The first paper to apply VQMC to a solid of nonlight elements was presented by 
Fahy, Wang, and Louie\cite{Fahy} 
dealing with the ground-state determination of 
diamond, graphite, and silicon. They proved that such a calculation is feasible 
if one relies on a pseudopotential to replace the inner shells' contribution. 
Different from the low atomic number systems, 
\cite{CeperleyAlder87,SugiyamaAlder,EcksteinSchattke} 
an uncertainty thereby enters into this technique by the construction of the 
pseudopotential itself. In order to treat the system 
on an equal footing with LDA, it is adequate to use the 
nonlocal {\it ab initio} potential developed by Hamann.\cite{Hamann} 

In this paper we apply the VQMC method to the compound semiconductor 
GaAs to calculate the ground-state properties. 
We closely rely on our former paper on lithium \cite{EcksteinSchattke} 
in treating the solid 
by simulating a finite system and in treating various other calculational 
details. Atomic simulations are carried through to obtain the binding 
energy with respect to these reference values. 
Hartree-Fock calculations have been peformed to extract the 
correlation energy. The inner shell electrons are included within a 
nonlocal pseudopotential for the valence electrons.

In Sec.~\ref{method} we briefly expose the theoretical method used by 
VQMC. The results are discussed in Sec.~\ref{results}.

\section{Method}
\label{method}
The Monte Carlo algorithm calculates expectation values of the form 
\begin{eqnarray}
  \langle f\rangle & = & \frac{\int f(x) p(x) dx}{\int p(x) dx} \text{,}
\end{eqnarray}
where the integration variable $x$ denotes typically 
several hundred single variables. 
The error is of purely statistical nature and depends 
on the number of simulation points as well as on the variance and the 
autocorrelation function of $f$.
The quantum mechanical expectation value of an observable $B$ 
is calculated by setting 
\begin{eqnarray}
  p = |\psi|^2 & \mbox{,\qquad} & 
  f = (B\psi) / \psi \text{.}
\end{eqnarray}
The smaller the variance of $(B\psi) / \psi$, the smaller the statistical 
error of the calculation, which can be taken as a measure for the 
many-body wave function $\psi$ to be an eigenfunction of $B$.

In the present approach we simulated $N$ valence electrons in a supercell 
containing a limited number of crystal unit cells. 
We sampled systems of 4 and 32 GaAs unit cells to test for convergence. 
Only the results for the large system with 256 electrons are described here. 
To model the solid, periodic boundary conditions are 
assumed, which affect the Hamiltonian and the wave function as well.
The Hamiltonian
\begin{eqnarray}
\label{Ham}
  H & = & T + E_c + E_{\rm loc} + E_{\rm nloc} 
\end{eqnarray}
consists of the kinetic energy $T$, the Coulomb energy $E_c$, and 
the local part $E_{\rm loc}$ as well as the nonlocal part $E_{\rm nloc}$ 
of the additional electron-ion pseudopotential, 
which is a generalized norm-conserving 
{\it ab initio} atomic pseudopotential.\cite{Hamann} 
The Coulomb energy includes the nucleon-nucleon, the electron-ion, and 
the electron-electron part.
It is calculated by means of the Ewald summation technique.\cite{Sugiyama} 
The local part of the pseudopotential 
can be decomposed into a sum over atoms $\alpha$ at site ${\bf R}_\alpha$,
\begin{eqnarray}
\label{locPP}
  E_{\rm loc} & = & \sum_{\alpha , {\bf R}_\alpha , s,i} 
            V_{{\rm loc},\alpha}({\bf r}_i^s)
\end{eqnarray}
with ${\bf r}_i^s$ being the position relative to ${\bf R}_\alpha$
of the $i$th electron of spin $s$. 
The respective sum for 
the nonlocal % semilocal 
pseudopotential contains a summation over angular momenta $l$, 
upon which the wave function is projected: \\
%\end{minipage}
%\\
%\begin{minipage}[t]{\textwidth}
\begin{eqnarray}
\label{nonlocPP}
  V_{{\rm nloc},\alpha}({\bf r}_i^s={\bf r}) & = & 
    \sum_{l} V_{\alpha,l}(r) 
    \frac{Y_{l0}(0,0) \int_{r'=r} Y^*_{l0}(\Omega_{r^\prime})
          \psi(\ldots,{\bf r}_i^s={\bf r}^\prime,\ldots) d\Omega_{r^\prime}}
         {\psi(\ldots,{\bf r}_i^s={\bf r},\ldots)} \text{.}
\end{eqnarray}
%\end{minipage}
%\\
%\begin{minipage}[t]{\textwidth}
%\twocolumn
The spherical angle $\Omega_{r^\prime}$ is referred to the direction of 
${\bf r}$ in an arbitrary but fixed polar coordinate system.
The projection operator demanding high numerical efforts 
needs only to be calculated 
where the radial pseudopotential $V_{\alpha,l}(r_i^s)$ is nonzero, 
i.e., when ${\bf r}_i^s$ lies within the radius of the atomic 
pseudopotential. The integration is done as a sum over a grid of four points 
on the sphere that is exact for $l\leq2$.\cite{Lebedev}

The wave function consists of a product of two Slater determinants and a 
Jastrow factor: 
\begin{eqnarray}
  \Psi({\bf r}_1,\ldots , {\bf r}_N) & = & 
         D^{\uparrow}({\bf r}_1,\ldots,{\bf r}_{N/2}) 
         D^{\downarrow}({\bf r}_{N/2+1},\ldots,{\bf r}_N)
%         \nonumber \\ & & \times
         \exp\left(-\sum_{i<j} u(r_{ij}) \right) \text{.}
\end{eqnarray}
The exponent of the Jastrow factor is chosen in the usual form for solids: 
\begin{eqnarray}
\label{Jasfac}
  u(r) & = & \frac{A}{r}(1-e^{-r/F}) \text{,}
\end{eqnarray}
where $A$ is a variational parameter. For a given $A$ the value of $F$ is 
fixed by the cusp condition, which removes the Coulomb singularity. 
The Slater determinants $D^{\uparrow},D^{\downarrow}$ consist of 
parametrized hybrid bonds % $\phi_{\vec \tau}$ 
in the tetrahedral directions ${\vec \tau}$ 
from the arsenic ion
\begin{eqnarray}
\label{SlaForm}
  \phi({\bf r}) & = & 
     \phi_{\rm hyb}^{\rm As}
   + \beta \phi_{\rm hyb}^{\rm Ga} 
  \text{,} \\
  \phi_{\rm hyb}^{\rm As}({\bf r}) & = & 
    \gamma^{\rm As} \phi_{s}^{\rm As}[({\bf r}-{\bf R}_{\rm As})/
    \zeta_{s}^{\rm As}]
   + \sum_i {\tau}_i {\phi}_{p_i}^{\rm As} 
           [({\bf r}-{\bf R}_{\rm As})/\zeta_{p}^{\rm As}] 
  \text{,} \\
  \phi_{\rm hyb}^{\rm Ga}({\bf r}) & = & 
     \gamma^{\rm Ga} \phi_{s}^{\rm Ga}
        [({\bf r}-{\bf R}_{\rm Ga})/\zeta_{s}^{\rm Ga}]
%  \nonumber \\ & & 
   - \sum_i {\tau}_i {\phi}_{p_i}^{\rm Ga}
        [({\bf r}-{\bf R}_{\rm Ga})/\zeta_{p}^{\rm Ga}] \text{,}
\end{eqnarray}
where $\beta$,
$\gamma$, %  $\gamma^{\rm As}$, $\gamma^{\rm Ga}$,
and $\zeta$ 
%  \zeta_{s}^{\rm As},\zeta_{p}^{\rm As},\zeta_{s}^{\rm Ga},\zeta_{p}^{\rm Ga}$ 
are variational parameters and $i$ runs over the Cartesian components 
$(x,y,z)$.
The quantities $\zeta$ describe the contraction of the wave functions 
as usual for atomic orbitals in solids and reduce the additional extent 
introduced by the Jastrow factor.
The coefficients $\gamma$ and $\beta$ represent the adjustable parameters for 
building hybrid bonds. The wave function is automatically normalized within 
the QMC formalism.

For the ratios $\Psi_{\rm new}/\Psi_{\rm old}$ of the random walk 
and in the projection operator 
as well as in the parts contributing to the kinetic energy, 
the inverse matrices 
of the Slater matrices are conveniently determined numerically by 
the Sherman-Morrison-Woodbury formula.% \cite{SherMorWood} 
%This formula yields the resulting 
%inverse of a matrix where one column has been changed, 
%when one electron, say $i$, is moved.
%\begin{eqnarray*}
% q := \frac{\det ( D_{lm}^s ({\rm new}))}{\det ( D_{lm}^s ({\rm old}))}
% & = &
%  \sum_{k=1}^{n} \overline{D}_{ki}^s ({\rm old}) D_{ki}^s ({\rm new}) \\
% D_{ki}^s ({\rm new}) & = & \Phi_{k,s}({\bf r}_i^s({\rm new}))
%\end{eqnarray*}
%\begin{eqnarray*}
%  \overline{D}_{lm}^s({\rm new}) & = & 
%  \left \lbrace \begin{array}{ll} 
%    \frac{1}{q} \overline{D}_{lm}^s({\rm old})
%   & \text{for $m = i$} \\
%   \overline{D}_{lm}^s({\rm old}) 
%    - \frac1q \overline{D}_{li}^s({\rm old})
%     \sum \limits_{k=1}^{n} 
%     \overline{D}_{km}^s({\rm old}) D_{ki}^s({\rm new})
%   & \text{for $m\neq i$}
%  \end{array} \right.
%\end{eqnarray*}
%\begin{eqnarray*}
%  \frac{{\rm \nabla}_{i} \det (D_{lm}^{s})}{\det (D_{lm}^{s})} & = & 
%  \sum_{k} 
%  ({\rm \nabla} \Phi_{k,s})({\rm r}_i^s) \overline{D}_{ki}^{s} \text{,} \\
%  \frac{\Delta_{i} \det (D_{lm}^{s})}{\det (D_{lm}^{s})} & = & 
%  \sum_{k} (\Delta \Phi_{k,s})({\rm r}_i^s) \overline{D}_{ki}^{s} \text{.}
%\end{eqnarray*}
The periodic boundary conditions with respect to the wave functions are 
controlled by Green's relation 
$-\int \Psi^* \Delta \Psi = \int |\nabla \Psi|^2$.

The minimization over the huge parameter space is done as follows.
Several hundreds of parameter sets in the region of the guessed minimum 
are chosen from a multidimensional Gaussian distribution. 
The total-energy expectation values 
weighted with their reciprocal statistical errors 
are fit into a multidimensional quadratic function. If the minimum of the fit 
lies within the region of the chosen statistical ensemble, 
a few parameter sets are selected along the minimum 
line of the fit with an increased number of simulation points.
This calculation along the minimum line is statistically independent of 
the prior calculation with the large number of parameter sets and therefore not 
biased as it would be, if the absolute minima of several calculations were 
explicitly taken.

\section{Results}
\label{results}
\begin{figure*}
 \psfig{file=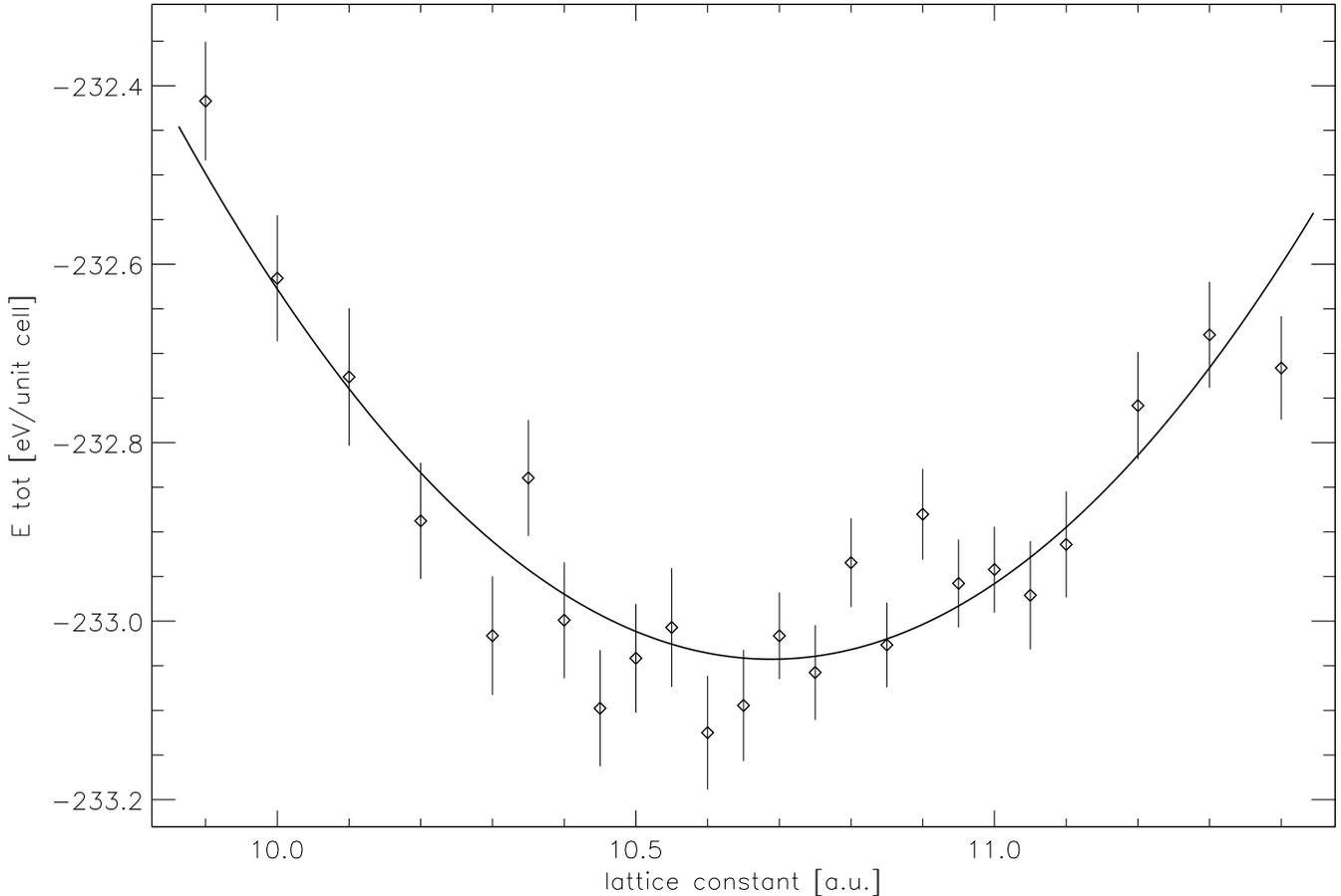,width=17.8cm}
% \vspace*{5.2in}
%\vspace*{2in}
\caption{Total energy vs lattice constant of gallium arsenide from a
VQMC calculation including 256 valence electrons; 
the curve is a quadratic fit.}
\label{Etot256}
\end{figure*}

At first, a minimization for the ground-state energies of the gallium and the 
arsenic atom has been carried through. Table \ref{atoms}
gives an overview of the obtained quantities. 
The total energies are lower than the experimental values, which are the sums 
of the ionization potentials of the valence electrons. 
Only the valence electrons have been simulated in this approach. 
The differences are supposed to indicate the error from the frozen-core 
approximation using 
pseudopotentials, also apparent in the first ionization potentials. 
As expected, the $\zeta$ parameters 
all result in $1.0$ from the minimization of the Hartree-Fock approach.
Introducing the Jastrow factor we observe compression of the atomic orbitals
even in the free atom.
To deduce the constituents of the total energy, 
we made a Hartree-Fock simulation.
The first ionization potentials are obtained from the approximation 
that the wave function is represented by the same parametrization scheme, 
i.e., by that of the neutral atom.
%\end{minipage}
%\\
%\begin{minipage}[t]{16cm}
% \onecolumn
\begin{table}
\caption{
Ground-state energies in eV 
of the gallium and arsenic atom from 
a variational QMC calculation: 
the first two lines are from Hartree-Fock calculations, 
the Hartree part of the electron-electron interaction is externally calculated, 
the exchange energy is the difference between 
the Hartree-Fock interaction energy and that Hartree part, 
the correlation energy is the difference between 
the total energy and the Hartree-Fock minimum total energy;
the experimental total energy is the sum over the ionization potentials 
of the simulated electrons;
the last line gives the first ionization potential.
}
\label{atoms}
  \begin{tabular}{l d r@{\,$\pm$}l d r@{\,$\pm$}l}
   \multicolumn{1}{l}{Energy}
   & \multicolumn{1}{c}{Ga expt.} & \multicolumn{2}{c}{Ga VQMC} 
   & \multicolumn{1}{c}{As expt.} & \multicolumn{2}{c}{As VQMC} \\
   \hline \hline
   Hartree-Fock, kinetic & & 20.28 & 0.01 & & 54.79 & 0.03 \\
   Hartree-Fock, potential & & $-$77.04 & 0.01 & & $-$222.92 & 0.04 \\
   Hartree, electron-electron & & 42.08 & 0.04 & & 134.3 & 0.7 \\
   Exchange, electron-electron & & $-$16.21 & 0.04 & & $-$37.3 & 0.8 \\
   Correlation, electron-electron & & $-$1.250 & 0.008 & & $-$2.00 & 0.02 \\
   Total & $-$57.21 & $-$58.000 & 0.003 & 
                      $-$169.554 & $-$170.134 & 0.007 \\
   Ionization & 5.999 & 6.10 & 0.02 & 9.81 & 10.38 & 0.02 
  \end{tabular}
\end{table}
%\nopagebreak
%
%\end{minipage}
%\\
%\begin{minipage}[t]{\textwidth}
%\twocolumn
\begin{table*}
\caption{Ground-state energies in eV per unit cell of gallium arsenide 
from a variational QMC calculation; 
$E_{\rm kin}$ and $E_{\rm pot}$ are produced by Hartree-Fock calculations, 
$E_{\rm corr}$ is the difference between 
$E_{\rm tot}$ and the Hartree-Fock minimum total energy,
$E_{\rm coh}$ is the difference between 
$E_{\rm tot}$ and the atomic energies, 
the experimental cohesive energy has been taken from 
Ref. \protect\onlinecite{Arthur}, 
the experimental total energy is the sum of the cohesive energy and 
the atomic total energies; 
$a_0$ denotes the lattice constant of the zinc-blende 
crystal and $B$ is the bulk modulus at temperature $T=0$;
LDA as well as the Becke and the Perdew gradient corrections (BP) 
are from density functional calculations of 
Ref. \protect\onlinecite{Ortiz}.
}
\label{GaAs}
  \begin{tabular}{l r@{\,}l r@{\,$\pm$}l r r}
     \multicolumn{1}{c}{GaAs}
   & \multicolumn{2}{c}{Expt.} & \multicolumn{2}{c}{VQMC} 
   & \multicolumn{1}{c}{LDA} & \multicolumn{1}{c}{BP} \\
   \hline \hline
   $E_{\rm kin}$ & & & 84.6 & 0.2 & & \\
   $E_{\rm pot}$ & & & $-$311.3 & 0.3 & & \\
   $E_{\rm corr}$ & & & $-$6.42 & 0.2 & & \\
   $E_{\rm tot}$ & $-$233.43 & $\pm$0.09 & $-$233.04 & 0.08 & & \\
   $E_{\rm coh}$ & $-$6.67 & $\pm$0.09 & $-$4.9 & 0.2 & $-$8.16 & $-$6.45 \\
   $a_0$ in a.u. & 10.6830 & & 10.69 & 0.1 & 10.41 & 10.70 \\
   $B$ in kbar & 756 & & 786 & 100 & & 
  \end{tabular}
\end{table*}
Figure \ref{Etot256}
shows the ground-state total energy of the gallium arsenide solid 
with $256$~electrons as a function of the lattice constant,
minimized with respect to all wave-function parameters. 
The curve is the quadratic function  
obtained from a weighted Gaussian fit of the data points.
Table \ref{GaAs}
shows the obtained quantities in comparison with the experimental ones 
and density functional calculations.\cite{Ortiz}
The best agreement with the experiment is found for the total energy. 
The pseudopotentials give excellent results in this approach of 
parametrized linear combination of atomic orbitals 
wave functions even with the semilocal approximation. 
Differences in the cohesive energy originate from underestimating the 
atomic total energies of As and Ga owing to deficiencies of the pseudopotential 
in the atomic calculation. 
The lattice constant agrees very well with the experimental value, 
although this quantity statistically arises with a somewhat larger 
uncertainty than the total energy. 
The respective error given in Table \ref{GaAs} is an upper bound 
estimated by just considering the fit in Fig.~\ref{Etot256}.
%\end{minipage}
%\\
%\begin{minipage}[t]{\textwidth}
% \onecolumn
%\end{minipage}
%\begin{minipage}[t]{\textwidth}
%\twocolumn
The bulk modulus is obtained slightly too high, which indicates that 
lattice constants in the far neighborhood of the minimum are treated 
worse in the minimization procedure than those in the close neighborhood.
This may be understood by observing that the minimum line is the result of 
the multidimensional quadratic fit, 
i.e., a straight line through the parameter space. In the far neighborhood 
of the global minimum this restriction could yield nonminimum parameters. 
The parts the total energy is composed of, as given in Table \ref{GaAs},
are calculated in the minimum of the Hartree-Fock approach.
The minimization was more difficult in this case, because 
the Coulomb cusp leads to a larger variance in the total energy.
The difference between the results of the
Hartree-Fock ansatz and the Jastrow-Slater ansatz, 
which defines the correlation energy, 
shows no statistically significant behavior in the simulated range 
of the lattice constant. 
Therefore the value of 
$-6.42$~eV per unit cell does not vary much near the ground-state density. 
The statistical data set for the correlation yields a straight 
regression line, the energy increasing with decreasing electron density 
with a slope of
$0.9\pm0.9$~eV per $r_s$. A measure for the ionicity is obtained by counting 
the electron visits to specific Voronoi polyhedra during a run. 
The polyhedra are constructed according to a
partition of the crystal in which lattice sites are added 
at the empty tetrahedral positions and at the cube center, the so-called 
empty spheres' positions in zinc-blende muffin-tin calculations. 
As a result the occupied As, occupied Ga, unoccupied As, 
and unoccupied Ga sites show 
charges of 3.9, 2.2, 0.9, and 1.0 electrons, respectively. Thus, both 
ions are positively charged the separated electrons extending into 
the interstitial space. The covalent bond charge was estimated by 
placing the centers of Voronoi polyhedra upon the bonds of the above lattice 
so that the polyhedra cover 
the bond region between sites that are occupied or empty. A slight charge 
accumulation appears on the occupied Ga-As bond of 0.13 electron per bond 
in addition to the 
charge which arises by the spherically symmetric electron distribution of the 
neutral atoms.

Summarizing the obtained results, the good agreement with experiment puts the 
variational quantum Monte Carlo method into competition with 
current density functional techniques as far as the 
absolute ground-state quantities, calculated here, are concerned. 
The minimum principle guarantees an 
additional reliability absent in DFT, which in contrast bears an uncertainty  
because of the, in some respects, arbitrary choice of the density functional. 
As the latter is able to treat more complicated solids, the comparison 
with QMC calculations on the basis of simple systems may lead to a 
selection and adjustment of suitable functionals.

\section*{ACKNOWLEDGMENTS}
We are indebted to Dr.~E.~Pehlke for various discussions on the topic of DFT.
The work was supported by the Deutsche Forschungsgemeinschaft under grants 
No.~Re~882/6-1 and No.~Scha~360/8-1.

%\end{minipage}

\end{document}